\documentclass[a4paper,11pt]{article}

\usepackage[utf8]{inputenc}
\usepackage[T1]{fontenc}
\usepackage{lmodern}
\usepackage{geometry}
\usepackage{graphicx}
\usepackage{authblk}
\usepackage{url}
\usepackage{cite}
\usepackage{fancyhdr}
\usepackage[hidelinks]{hyperref}
\usepackage{orcidlink}

\usepackage{amsmath}
\usepackage{amssymb}
\usepackage{subcaption}
\usepackage{tikz}
\usetikzlibrary{shapes,arrows,positioning,automata,backgrounds,calc,er,patterns}
\usepackage[compat=1.1.0]{tikz-feynman}

\fancypagestyle{firstpage}{
	\fancyhf{}
	\fancyhead[C]{\includegraphics[height=2.2cm]{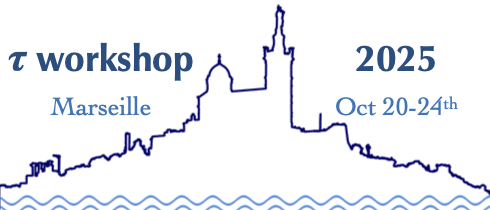}\\ \textit{Proceedings of the 18th International Workshop on Tau Lepton Physics}
	} % <-- put your logo here
}

\title{Flavour and precision probes of a class of scotogenic models }
\author[1]{A. Darricau\,\orcidlink{0009-0001-2953-7499}\thanks{Presenting author.}, H. Lee\,\orcidlink{0000-0002-7197-2937}}
\author[1]{J. Orloff\,\orcidlink{0000-0002-8482-0748}}
\author[1]{A. M. Teixeira\,\orcidlink{0000-0002-4184-2728}}

\affil[1]{Laboratoire de Physique de Clermont Auvergne (UMR 6533), CNRS/IN2P3, Univ. Clermont Auvergne, 4 Av. Blaise Pascal, 63178 Aubi\`ere Cedex, France}

\date{}

\begin{document}

\maketitle
\thispagestyle{firstpage}

%%%% if you speak on behalf of a collaboration
%\vspace{-1cm}
%\begin{center}
%	\textit{On behalf of the Example Collaboration}
%\end{center}
%\vspace{0.5cm}
%%%%%%%%%% if not comment

\begin{abstract}
	We address the phenomenological impact of a well-motivated class of scotogenic models regarding flavour and electroweak precision observables. For the case of a ``T1-2-A'' variant, we carry out a full computation of the next-to-leading order corrections to leptonic Higgs and $Z$-boson decays. 
    We revisit previously drawn conclusions on operator dominance and constraints on the parameter space in view of the evolution regarding $\Delta a_\mu$. 
    Finally, we consider the role of $H \to \mu\mu$ decays (and other flavour conserving Higgs decays), as well as precision observables in probing this class of models at future colliders.
\end{abstract}

\section{Introduction}

Despite its many successes, the Standard Model (SM) fails to explain several observations: neutrino oscillations, the baryon asymmetry of the Universe (BAU) and its abundance of cold dark matter (DM).

Many well motivated extensions of the SM have been proposed to address these experimental caveats while also trying to ease theoretical issues. Among these new physics (NP) proposals, ``scotogenic'' models are especially appealing, as they generate neutrino masses radiatively (at loop level) while providing a viable DM candidate stabilised by a discrete symmetry.
Departing from the original minimal model~\cite{Tao:1996vb,Ma:2006km,Fraser:2014yha}, numerous variants have been proposed - most aiming at enhancing the testability potential (for instance via charged lepton flavour violation (cLFV)~\cite{Toma:2013zsa,Vicente:2014wga,Rocha-Moran:2016enp,Avila:2019hhv,Baumholzer:2019twf,Ahriche:2020pwq,DeRomeri:2021yjo,Boruah:2021ayj,Liu:2022byu} or lepton number violation (LNV)~\cite{Mandal:2019oth,Bonilla:2019ipe,Ma:2021eko,DeRomeri:2022cem}), further considering the prospects for leptognesis.
A recent variant - the so-called ``T1-2-A'' setup - extends the SM via 1 (2) $SU(2)_L$ doublet and 1 (2) singlet scalar (fermions) to potentially address neutrino masses, DM, BAU through leptogenesis and the discrepancy in the anomalous magnetic moment of the muon ($\Delta a_\mu$)~\cite{Alvarez:2023dzz}. 
Notice however that recently, the formerly significant tension in $\Delta a_\mu$ (around  $5 \sigma$, relying on a data-driven computation of the hadronic vacuum polarisation contribution)~\cite{Aoyama:2020ynm}
% However, the tension of $5 \sigma$ in $\Delta a_\mu$, using the SM prediction proposed in the ``White paper''~\cite{Aoyama:2020ynm} (relying on a data-driven computation of the hadronic vacuum polarisation contribution), 
has been dramatically reduced to around $1\sigma$ in view of the evolution of first-principle computations of the hadronic contributions\cite{Budapest-Marseille-Wuppertal:2017okr,RBC:2018dos,Giusti:2019xct,Shintani:2019wai,FermilabLattice:2019ugu,Gerardin:2019rua,Borsanyi:2020mff,Lehner:2020crt,Aubin:2022hgm,Boccaletti:2024guq}.

In~\cite{Darricau:2025vcs}, a thorough study of cLFV and electroweak precision observables (EWPO) was carried out, taking into account the recent evolution in $\Delta a_\mu$;  particular emphasis was given to the role of EWPO, which can provide significant constraints to NP models featuring extensions of the lepton sector~\cite{Abada:2023raf} - especially given the high precision of current and future measurements (as in the case of FCC-ee, in its $Z$-pole run).

\newpage
\section{Brief description of the model}

We consider the model first explored in~\cite{Alvarez:2023dzz}, which  introduces new fields as summarised in Table~\ref{table:NPcontent}.
\begin{table}[h!]
\centering\renewcommand{\arraystretch}{1.4} 
\begin{tabular}{c|cc|cccc}
\hline
\hline
Field & $\eta$ & $S$ & $F_{1}$ & $F_{2}$ & $\Psi_{1}$ & $\Psi_{2}$\\ 
\hline
SU$(2)_L$ & $\mathbf{2}$ & $\mathbf{1}$ & $\mathbf{1}$ & $\mathbf{1}$ & $\mathbf{2}$ & $\mathbf{2}$ \\
U$(1)_Y$ & $1$ & $0$ & $0$ & $0$ & $-1$ & $1$  \\
\hline
\hline
\end{tabular}
\caption{Additional field content of the ``T1-2-A'' scotogenic model variant (cf.~\cite{Alvarez:2023dzz}). All fields are odd under the new $Z_2$ symmetry.} 
\label{table:NPcontent}
\renewcommand{\arraystretch}{1} 
\end{table}

\begin{figure}[h!]
    \centering
    \raisebox{8mm}{
    \begin{subfigure}[b]{0.4\textwidth}
        \centering
        \begin{tikzpicture}
        \begin{feynman}
        \vertex (a) at (0,0) {\(\nu_\alpha\)};
        \vertex (b) at (1,0);
        \vertex (c) at (2,0);
        \vertex (d) at (2,1);
        \vertex (e) at (3,0);
        \vertex (f) at (4,0) {\(\nu_\beta\)};
        \vertex (g) at (3,2) {\( H\)};
        \vertex (h) at (1,2) {\( H\)};
        \diagram* {
        (a) -- [fermion] (b),
        (b) -- [plain, edge label'=\( F_i\)] (c),
        (c) -- [plain, edge label'=\( F_i\)] (e),
        (b) -- [scalar, quarter left, edge label=\( \eta\)] (d),
        (d) -- [scalar, quarter left, edge label=\( \eta\)] (e),
        (b) -- [insertion = 0.5, edge label=\( M_{F_{ii}}\)] (e),
        (f) -- [fermion] (e),
        (d) -- [scalar] (g),
        (d) -- [scalar] (h)
        };
        \end{feynman}
        \end{tikzpicture}
    \end{subfigure}}
    \begin{subfigure}[b]{0.35\textwidth}
        \centering
        \begin{tikzpicture}
        \begin{feynman}
        \vertex (a) at (0,0) {\(\nu_\alpha\)};
        \vertex (b) at (1,0);
        \vertex (c) at (2,-1);
        \vertex (d) at (2,1);
        \vertex (e) at (3,0);
        \vertex (f) at (4,0) {\(\nu_\beta\)};
        \vertex (g) at (2,2) {\( H^\dagger\)};
        \vertex (h) at (2,-2) {\( H^\dagger\)};
        \diagram* {
        (a) -- [fermion] (b),
        (b) -- [plain, quarter right, edge label'=\( F_i\)] (c),
        (c) -- [plain, quarter right, edge label'=\( \psi_2\)] (e),
        (b) -- [scalar, quarter left, edge label=\( \eta\)] (d),
        (d) -- [scalar, quarter left, edge label=\( S\)] (e),
        (f) -- [fermion] (e),
        (d) -- [scalar] (g),
        (c) -- [scalar] (h)
        };
        \end{feynman}
        \end{tikzpicture}
    \end{subfigure}
    \caption{One-loop diagrams contributing to neutrino masses (in the interaction basis).}
    \label{fig:NMassInt}
\end{figure}
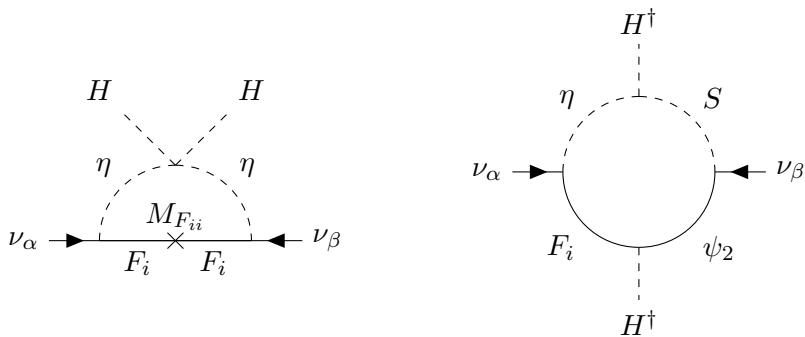

\noindent 
The SM interaction Lagrangian is generalised as follows, 
%additional terms now encode the interactions of the NP fields with the SM leptons, gauge bosons and the Higgs:
\begin{align}\label{eq:lagrangian}
\mathcal{L}_\text{fermion}\, &= i\, (\overline{\psi_i} \,\gamma^\mu\, D_\mu\, \psi_i + \overline{F_i} \,\gamma^\mu \,D_\mu \,F_i) \nonumber \\ 
&- M_\psi \,\bar{\psi_1} \,\tilde{\psi_2} - \frac{1}{2} \,{M_F}_{ii} \,\overline{F_i^c}\, F_i + y_{1 i}^*\, \overline{F_i} \,\Phi^\dagger \,\tilde{\psi_1} + y_{2 i}^* \,\overline{F_i} \,\Phi \,\psi_2^c \nonumber\\
&- g_\psi^\alpha \,\tilde{\bar{\psi_2}} \,L^\alpha_L S - g_{F_i}^\alpha \,\widetilde{\overline{L_L^\alpha}}\, \eta F_i - g_R^\alpha \overline{e_R^\alpha} \,\eta^\dagger \,\Psi_1 + \text{H.c.}\, ,\\
\mathcal{V}_\text{scalar}\, &= \frac{1}{2} M_S^2\, S^2 + \frac{1}{2} \lambda_{4 S} \,S^4 + M_\eta^2 \,|\eta|^2 + \lambda_{4 \eta} \,|\eta|^4 + \frac{1}{2} \lambda_S \,S^2 |H|^2 + \frac{1}{2} \lambda_{S \eta} \,S^2 |\eta|^2 \nonumber \\ &
+ \lambda_{\eta} \,|\eta|^2\,|H|^2 + \lambda_{\eta}^\prime \,|\eta H^\dagger|^2 + \frac{1}{2} \lambda_{\eta}^{\prime \prime} \left[ \left(H \eta^\dagger \right)^2 +\text{H.c.} \right] + \alpha \,S \left[ H \eta^\dagger + \text{H.c.} \right].
\end{align}
As common to all scotogenic realisations, the discrete $Z_2$ symmetry ensuring the stability of the lightest neutral NP particle precludes neutrino mass generation at the tree-level; loop-level generation (cf. Fig.~\ref{fig:NMassInt}), together with the smallness of the new couplings thus allows for a more natural explanation of neutrino masses and oscillation data in general.
In turn, this gives rise to contributions to the neutrino masses of the form $\overline{\nu_{\beta}^{c}} \left( \mathcal{M}_{\nu} \right)_{\beta \alpha} \nu_{\alpha}$; the neutrino mass matrix can be cast in a compact manner relying on a generalised matrix of ``couplings'' $\mathcal{G}$, and on the contributions arising from the exchange of the new massive fields in the loop,
$\mathcal{M}_{L}$~\cite{Alvarez:2023dzz}:
\begin{equation}\label{eq:Gmatrix}
    \mathcal{M}_{\nu} = \mathcal{G}^T \mathcal{M}_{L} \mathcal{G}, \quad \text{where} \quad \mathcal{G} = \begin{pmatrix}
        g_{\psi}^{e} & g_{\psi}^{\mu} & g_{\psi}^{\tau} \\[5pt]
        g_{F_{1}}^{e} & g_{F_{1}}^{\mu} & g_{F_{1}}^{\tau} \\[5pt]
        g_{F_{2}}^{e} & g_{F_{2}}^{\mu} & g_{F_{2}}^{\tau}
    \end{pmatrix}\,.
\end{equation}
%The naturally suppressed $\mathcal{M}_{L}$ allows for 
Sizable cLFV couplings are encoded in $\mathcal{G}$, 
%contributing to the active neutrino masses, 
leading to a rich phenomenology.
%in cLFV decays.
%
The model puts forward several 
% The $Z_2$ symmetry also constrains the lightest state of the extended fermion and scalar sector to be stable making them potential 
DM candidates, including the lightest CP-even scalar $\phi_{1}$, the CP-odd scalar $A^{0}$ and the lightest fermion $\chi_{1}^{0}$, provided that they lead to a viable 
% In order for the lightest stable neutral particle to be a potential viable DM candidate, it must give a viable 
relic density~\cite{Planck:2018vyg} and comply with constraints arising from numerous direct and indirect DM searches.
To do so, we rely on {micrOMEGAs}~\cite{Alguero:2023zol}, which provides a full evaluation of the dark matter prospects of a given NP model.

\section{Flavour and electroweak observables in the scotogenic ``T1-2-A'' variant}

\begin{figure}[h!]
\centering
\includegraphics[width=0.42 \textwidth] {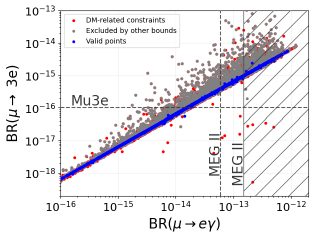}
\hspace*{10mm}
\includegraphics[width=0.42 \textwidth] {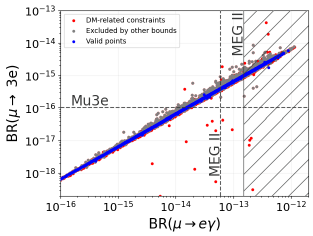}
\\
\includegraphics[width=0.42 \textwidth]{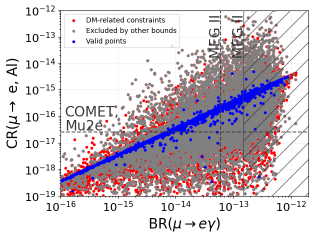}
\hspace*{10mm}
\includegraphics[width=0.42 \textwidth]{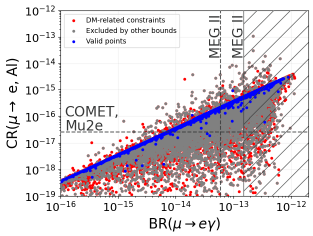}
\caption{Predicted rates for $\mu \to 3e$ (top) and $\mu -e$ conversion in nuclei (bottom), both versus BR($\mu \to e \gamma$). On the left panels, all displayed points lead to a SM-like $(g-2)_\mu$ (i.e. $\Delta a_\mu \approx 1.5\sigma$), while the right panels exhibit a significant NP contribution ($\Delta a_\mu \approx 4.2\sigma$).  
Red points denote exclusion due to conflict with dark matter constraints, grey points correspond to the violation of at least one phenomenological bound (flavour, EWPO) other than those under study; blue points are viable under all constraints but those under consideration. Full (dashed) lines correspond to current bounds (future sensitivity), with hatched areas already excluded.} 
\label{fig:Mu3e:Mu-e:mueg}
\end{figure}

As briefly mentioned in the Introduction, 
% the authors of~\cite{Alvarez:2023dzz} made use of the freedom of the couplings in the $\mathcal{G}$ matrix to saturate the previously existing $5 \sigma$ tension in $\Delta a_\mu$. With a 
the tension in $\Delta a_\mu$ has been 
dramatically reduced tension (to around $1 \sigma$), and as such noticeable changes are expected to occur in the phenomenology of cLFV decays. Following comprehensive scans of the parameter space, the updated prospects for cLFV observables are displayed in Fig.~\ref{fig:Mu3e:Mu-e:mueg}.
%In order to put forward these noticeable changes, comprehensive scans of the parameter space are conducted, and all available relevant constraints are applied - this include several cLFV, DM and EWPO related constraints.

The top plots of Fig.~\ref{fig:Mu3e:Mu-e:mueg} display 
the effect of reducing the tension in $\Delta a_\mu$ on the apparent correlation between BR($\mu \to 3 e$) and BR($\mu \to e \gamma$): a SM-like $\Delta a_\mu$ leads to a loss of correlation between the observables, albeit recovered when applying experimental constraints; the correlation has a strong experimental interest as the measurement of one cLFV observable inevitably predicts the observation of the other.
%On the top plots of Fig.~\ref{fig:Mu3e:Mu-e:mueg}, we display the effect of relaxing the tension in $\Delta a_\mu$ on the apparent correlation between BR($\mu \to 3 e$) and BR($\mu \to e \gamma$). This correlation is of particular interest as a measurement of BR($\mu \to e \gamma$) by future experiments would inevitably predict a measurement of BR($\mu \to 3 e$). We see from this top row that an SM-like $\Delta a_\mu$ leads to a loss of correlation between the observables, albeit recovered when applying experimental constraints. 
A similar effect can be seen in the bottom row for the cLFV $\mu - e$ conversion in nuclei.  
%and in $\tau \to 3 \mu$ decays, which can be explained by an enhancement 
These effects can be explained via the increase of the 
$Z$-Penguin contributions for a SM-like $\Delta a_\mu$ (compared to dominant $\gamma$-penguins for NP-like $\Delta a_\mu$).
%Although such enhancement could cause $Z$ and Higgs cLFV decays that were previously assumed to be minimal to become more relevant, predictions for these cLFV decays remain beyond hope of a measurement by future experiments.

In view of the potential of future lepton colliders in what concerns EW precision, addressing the impact of a given NP model regarding the latter becomes important. While invisible decay channels of the $Z$ and the Higgs constrain the NP states to be heavier than $m_{Z.H}/2$, corrections to flavour conserving $Z$ and Higgs decays offer non-trivial constraints.
%, as shown in Fig.~\ref{fig:ZMuMu}.
\begin{figure}[h!]
\centering
\includegraphics[width=0.42 \textwidth]{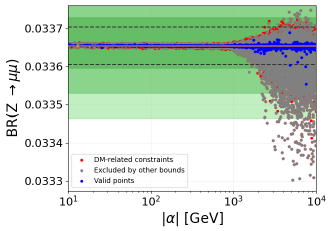}
\hspace*{10mm}
\includegraphics[width=0.42 \textwidth]
{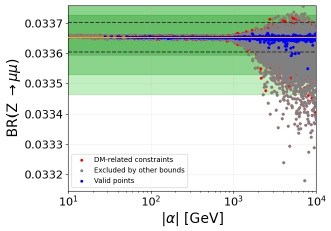}
\caption{BR($Z \to \mu \mu$) as a function of the scalar trilinear coupling $\alpha$. Point-colour scheme as in Fig.~\ref{fig:Mu3e:Mu-e:mueg}. From darker to lighter, the green bands denote current bounds at $1\sigma$, $2\sigma$ and $3\sigma$.
The full orange line corresponds to the SM 2-loop prediction~\cite{Dubovyk:2018rlg}, while dashed grey lines denote future sensitivity at FCC-ee.
%centred on the current experimental central value. 
The left (right) panel corresponds to a SM-like (NP-like) $(g-2)_\mu$.
}
\label{fig:ZMuMu}
\end{figure}

\begin{figure}[h!]
\centering
\includegraphics[width=0.42 \textwidth]{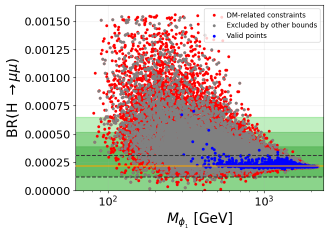}
\hspace*{10mm}
\includegraphics[width=0.42 \textwidth]
{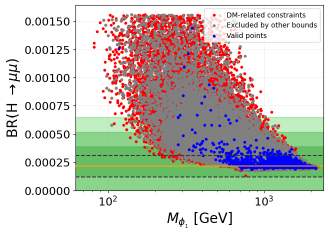}
\caption{BR($H \to \mu \mu$) as a function of the lightest scalar mass, $M_{\phi_1}$, for a SM-like $(g-2)_\mu$ (left) and a significant tension ($\Delta a_\mu = 4.2 \sigma$, on the right).
Line and colour code as in Fig.~\ref{fig:ZMuMu}, with the full line denoting the SM prediction~\cite{LHCHiggsCrossSectionWorkingGroup:2016ypw}.}
\label{fig:HMuMu}
\end{figure}
The results for corrections in $Z$ (Higgs) flavour conserving decays are presented in Fig.~\ref{fig:ZMuMu} (Fig.~\ref{fig:HMuMu}).  Regimes corresponding to a large trilinear coupling $\alpha \gtrsim 1$ TeV (or equivalently to a small mass for the lightest scalar $M_{\phi_1} \lesssim 1$ TeV) can lead to sizeable corrections already excluded by current experiments (especially $H \to \ell_\alpha \ell_\alpha$). Although relaxing the tension in $\Delta a_\mu$ mitigates the important corrections in the latter regimes, these remain largely disfavoured.
Concerning the impact for lepton flavour universality violation (LFUV), notice that  $\mathcal{G}$ matrix can also induce them, but the new effects remain too small to be measured.
%Investigating the ratios of these flavour conserving decays between different flavours can teach us about lepton flavour universality violation (LFUV) which the SM only predicts to be caused by mass effects therefore being small. Although LFUV now also happens through the $\mathcal{G}$ matrix, its effect remain too small to be measured.
Finally NP corrections to the $S$, $T$ and $U$ parameters were also investigated but played a non-constraining role on the parameter space (although corrections within future reach are possible).

\section{Conclusions}
The ``T1-2-A'' scotogenic framework remains theoretically appealing as it connects radiative neutrino mass generation with a viable dark matter candidate. In this work we have put forward the consequences of relaxing the tension in the muon $g-2$ on cLFV observables, putting forward the correlation between the observables rendering this model easily testable. We also investigated NP corrections to EWPO, leading to sizeable corrections to $Z$ and Higgs flavour conserving decays, making $Z/H \to \mu \mu, \tau \tau$ decays prime probes for this model.

\bibliographystyle{ieeetr}
\bibliography{references}

@article{Tao:1996vb,
    author = "Tao, Zhi-jian",
    title = "{Radiative seesaw mechanism at weak scale}",
    eprint = "hep-ph/9603309",
    archivePrefix = "arXiv",
    doi = "10.1103/PhysRevD.54.5693",
    journal = "Phys. Rev. D",
    volume = "54",
    pages = "5693--5697",
    year = "1996"
}

@article{Ma:2006km,
    author = "Ma, Ernest",
    title = "{Verifiable radiative seesaw mechanism of neutrino mass and dark matter}",
    eprint = "hep-ph/0601225",
    archivePrefix = "arXiv",
    reportNumber = "UCRHEP-T403",
    doi = "10.1103/PhysRevD.73.077301",
    journal = "Phys. Rev. D",
    volume = "73",
    pages = "077301",
    year = "2006"
}

@article{Fraser:2014yha,
    author = "Fraser, Sean and Ma, Ernest and Popov, Oleg",
    title = "{Scotogenic Inverse Seesaw Model of Neutrino Mass}",
    eprint = "1408.4785",
    archivePrefix = "arXiv",
    primaryClass = "hep-ph",
    reportNumber = "UCRHEP-T543-(AUG-2014)",
    doi = "10.1016/j.physletb.2014.08.069",
    journal = "Phys. Lett. B",
    volume = "737",
    pages = "280--282",
    year = "2014"
}

@article{Toma:2013zsa,
    author = "Toma, Takashi and Vicente, Avelino",
    title = "{Lepton Flavor Violation in the Scotogenic Model}",
    eprint = "1312.2840",
    archivePrefix = "arXiv",
    primaryClass = "hep-ph",
    reportNumber = "DCPT-13-198",
    doi = "10.1007/JHEP01(2014)160",
    journal = "JHEP",
    volume = "01",
    pages = "160",
    year = "2014"
}

@article{Vicente:2014wga,
    author = "Vicente, Avelino and Yaguna, Carlos E.",
    title = "{Probing the scotogenic model with lepton flavor violating processes}",
    eprint = "1412.2545",
    archivePrefix = "arXiv",
    primaryClass = "hep-ph",
    reportNumber = "MS-TP-14-37",
    doi = "10.1007/JHEP02(2015)144",
    journal = "JHEP",
    volume = "02",
    pages = "144",
    year = "2015"
}

@article{Rocha-Moran:2016enp,
    author = "Rocha-Moran, Paulina and Vicente, Avelino",
    title = "{Lepton Flavor Violation in the singlet-triplet scotogenic model}",
    eprint = "1605.01915",
    archivePrefix = "arXiv",
    primaryClass = "hep-ph",
    reportNumber = "IFIC-16-24, BONN-TH-2016-03",
    doi = "10.1007/JHEP07(2016)078",
    journal = "JHEP",
    volume = "07",
    pages = "078",
    year = "2016"
}

@article{Avila:2019hhv,
    author = "\'Avila, Ivania M. and De Romeri, Valentina and Duarte, Laura and Valle, Jos\'e W. F.",
    title = "{Phenomenology of scotogenic scalar dark matter}",
    eprint = "1910.08422",
    archivePrefix = "arXiv",
    primaryClass = "hep-ph",
    reportNumber = "IFIC/19-XXX",
    doi = "10.1140/epjc/s10052-020-08480-z",
    journal = "Eur. Phys. J. C",
    volume = "80",
    number = "10",
    pages = "908",
    year = "2020"
}

@article{Baumholzer:2019twf,
    author = "Baumholzer, Sven and Brdar, Vedran and Schwaller, Pedro and Segner, Alexander",
    title = "{Shining Light on the Scotogenic Model: Interplay of Colliders and Cosmology}",
    eprint = "1912.08215",
    archivePrefix = "arXiv",
    primaryClass = "hep-ph",
    reportNumber = "MITP/19-088",
    doi = "10.1007/JHEP09(2020)136",
    journal = "JHEP",
    volume = "09",
    pages = "136",
    year = "2020"
}

@article{Ahriche:2020pwq,
    author = "Ahriche, Amine and Jueid, Adil and Nasri, Salah",
    title = "{A natural scotogenic model for neutrino mass \& dark matter}",
    eprint = "2007.05845",
    archivePrefix = "arXiv",
    primaryClass = "hep-ph",
    doi = "10.1016/j.physletb.2021.136077",
    journal = "Phys. Lett. B",
    volume = "814",
    pages = "136077",
    year = "2021"
}

@article{DeRomeri:2021yjo,
    author = "De Romeri, Valentina and Puerta, Miguel and Vicente, Avelino",
    title = "{Dark matter in a charged variant of the Scotogenic model}",
    eprint = "2106.00481",
    archivePrefix = "arXiv",
    primaryClass = "hep-ph",
    reportNumber = "IFIC/21-19",
    doi = "10.1140/epjc/s10052-022-10532-5",
    journal = "Eur. Phys. J. C",
    volume = "82",
    number = "7",
    pages = "623",
    year = "2022"
}

@article{Boruah:2021ayj,
    author = "Boruah, Bichitra Bijay and Sarma, Lavina and Das, Mrinal Kumar",
    title = "{Lepton flavor violation and leptogenesis in discrete flavor symmetric scotogenic model}",
    doi = "10.1016/j.nuclphysb.2021.115472",
    journal = "Nucl. Phys. B",
    volume = "969",
    pages = "115472",
    year = "2021"
}

@article{Liu:2022byu,
    author = "Liu, Jiao and Han, Zhi-Long and Jin, Yi and Li, Honglei",
    title = "{Unraveling the Scotogenic model at muon collider}",
    eprint = "2207.07382",
    archivePrefix = "arXiv",
    primaryClass = "hep-ph",
    doi = "10.1007/JHEP12(2022)057",
    journal = "JHEP",
    volume = "12",
    pages = "057",
    year = "2022"
}

@article{Mandal:2019oth,
    author = "Mandal, Sanjoy and Rojas, Nicol\'as and Srivastava, Rahul and Valle, Jos\'e W. F.",
    title = "{Dark matter as the origin of neutrino mass in the inverse seesaw mechanism}",
    eprint = "1907.07728",
    archivePrefix = "arXiv",
    primaryClass = "hep-ph",
    reportNumber = "IFIC/19-XXX",
    doi = "10.1016/j.physletb.2021.136609",
    journal = "Phys. Lett. B",
    volume = "821",
    pages = "136609",
    year = "2021"
}

@article{Bonilla:2019ipe,
    author = "Bonilla, Cesar and de la Vega, Leon M. G. and Lamprea, J. M. and Lineros, Roberto A. and Peinado, Eduardo",
    title = "{Fermion Dark Matter and Radiative Neutrino Masses from Spontaneous Lepton Number Breaking}",
    eprint = "1908.04276",
    archivePrefix = "arXiv",
    primaryClass = "hep-ph",
    doi = "10.1088/1367-2630/ab7254",
    journal = "New J. Phys.",
    volume = "22",
    number = "3",
    pages = "033009",
    year = "2020"
}

@article{Ma:2021eko,
    author = "Ma, Ernest and De Romeri, Valentina",
    title = "{Radiative seesaw dark matter}",
    eprint = "2105.00552",
    archivePrefix = "arXiv",
    primaryClass = "hep-ph",
    doi = "10.1103/PhysRevD.104.055004",
    journal = "Phys. Rev. D",
    volume = "104",
    number = "5",
    pages = "055004",
    year = "2021"
}

@article{DeRomeri:2022cem,
    author = "De Romeri, Valentina and Nava, Jacopo and Puerta, Miguel and Vicente, Avelino",
    title = "{Dark matter in the scotogenic model with spontaneous lepton number violation}",
    eprint = "2210.07706",
    archivePrefix = "arXiv",
    primaryClass = "hep-ph",
    reportNumber = "IFIC/22-24",
    doi = "10.1103/PhysRevD.107.095019",
    journal = "Phys. Rev. D",
    volume = "107",
    number = "9",
    pages = "095019",
    year = "2023"
}

@article{Alvarez:2023dzz,
    author = "Alvarez, A. and Banik, A. and Cepedello, R. and Herrmann, B. and Porod, W. and Sarazin, M. and Schnelke, M.",
    title = "{Accommodating muon (g \ensuremath{-} 2) and leptogenesis in a scotogenic model}",
    eprint = "2301.08485",
    archivePrefix = "arXiv",
    primaryClass = "hep-ph",
    reportNumber = "LAPTH-060/22",
    doi = "10.1007/JHEP06(2023)163",
    journal = "JHEP",
    volume = "06",
    pages = "163",
    year = "2023"
}

@article{Abada:2023raf,
    author = "Abada, A. and Kriewald, J. and Pinsard, E. and Rosauro-Alcaraz, S. and Teixeira, A. M.",
    title = "{Heavy neutral lepton corrections to SM boson decays: lepton flavour universality violation in low-scale seesaw realisations}",
    eprint = "2307.02558",
    archivePrefix = "arXiv",
    primaryClass = "hep-ph",
    doi = "10.1140/epjc/s10052-023-12364-3",
    journal = "Eur. Phys. J. C",
    volume = "84",
    number = "2",
    pages = "149",
    year = "2024"
}

@article{Aoyama:2020ynm,
    author = "Aoyama, T. and others",
    title = "{The anomalous magnetic moment of the muon in the Standard Model}",
    eprint = "2006.04822",
    archivePrefix = "arXiv",
    primaryClass = "hep-ph",
    reportNumber = "FERMILAB-PUB-20-207-T, INT-PUB-20-021, KEK Preprint 2020-5,
  MITP/20-028, KEK Preprint 2020-5, MITP/20-028, CERN-TH-2020-075, IFT-UAM/CSIC-20-74, LMU-ASC 18/20, LTH 1234,
  LU TP 20-20, LTH 1234, LU TP 20-20, MAN/HEP/2020/003, PSI-PR-20-06, UWThPh 2020-14, ZU-TH 18/20",
    doi = "10.1016/j.physrep.2020.07.006",
    journal = "Phys. Rept.",
    volume = "887",
    pages = "1--166",
    year = "2020"
}

@article{Budapest-Marseille-Wuppertal:2017okr,
    author = "Borsanyi, Sz. and others",
    collaboration = "Budapest-Marseille-Wuppertal",
    title = "{Hadronic vacuum polarization contribution to the anomalous magnetic moments of leptons from first principles}",
    eprint = "1711.04980",
    archivePrefix = "arXiv",
    primaryClass = "hep-lat",
    doi = "10.1103/PhysRevLett.121.022002",
    journal = "Phys. Rev. Lett.",
    volume = "121",
    number = "2",
    pages = "022002",
    year = "2018"
}

@article{RBC:2018dos,
    author = {Blum, T. and Boyle, P. A. and G\"ulpers, V. and Izubuchi, T. and Jin, L. and Jung, C. and J\"uttner, A. and Lehner, C. and Portelli, A. and Tsang, J. T.},
    collaboration = "RBC, UKQCD",
    title = "{Calculation of the hadronic vacuum polarization contribution to the muon anomalous magnetic moment}",
    eprint = "1801.07224",
    archivePrefix = "arXiv",
    primaryClass = "hep-lat",
    doi = "10.1103/PhysRevLett.121.022003",
    journal = "Phys. Rev. Lett.",
    volume = "121",
    number = "2",
    pages = "022003",
    year = "2018"
}

@article{Giusti:2019xct,
    author = "Giusti, D. and Lubicz, V. and Martinelli, G. and Sanfilippo, F. and Simula, S.",
    title = "{Electromagnetic and strong isospin-breaking corrections to the muon $g - 2$ from Lattice QCD+QED}",
    eprint = "1901.10462",
    archivePrefix = "arXiv",
    primaryClass = "hep-lat",
    doi = "10.1103/PhysRevD.99.114502",
    journal = "Phys. Rev. D",
    volume = "99",
    number = "11",
    pages = "114502",
    year = "2019"
}

@article{Shintani:2019wai,
    author = "Shintani, Eigo and Kuramashi, Yoshinobu",
    collaboration = "PACS",
    title = "{Hadronic vacuum polarization contribution to the muon $g-2$ with 2+1 flavor lattice QCD on a larger than (10 fm$)^4$ lattice at the physical point}",
    eprint = "1902.00885",
    archivePrefix = "arXiv",
    primaryClass = "hep-lat",
    doi = "10.1103/PhysRevD.100.034517",
    journal = "Phys. Rev. D",
    volume = "100",
    number = "3",
    pages = "034517",
    year = "2019"
}

@article{FermilabLattice:2019ugu,
    author = "Davies, C. T. H. and others",
    collaboration = "Fermilab Lattice, LATTICE-HPQCD, MILC",
    title = "{Hadronic-vacuum-polarization contribution to the muon\textquoteright{}s anomalous magnetic moment from four-flavor lattice QCD}",
    eprint = "1902.04223",
    archivePrefix = "arXiv",
    primaryClass = "hep-lat",
    reportNumber = "FERMILAB-PUB-19-064-T",
    doi = "10.1103/PhysRevD.101.034512",
    journal = "Phys. Rev. D",
    volume = "101",
    number = "3",
    pages = "034512",
    year = "2020"
}

@article{Gerardin:2019rua,
    author = {G\'erardin, Antoine and C\`e, Marco and von Hippel, Georg and H\"orz, Ben and Meyer, Harvey B. and Mohler, Daniel and Ottnad, Konstantin and Wilhelm, Jonas and Wittig, Hartmut},
    title = "{The leading hadronic contribution to $(g-2)_\mu$ from lattice QCD with $N_{\rm f}=2+1$ flavours of O($a$) improved Wilson quarks}",
    eprint = "1904.03120",
    archivePrefix = "arXiv",
    primaryClass = "hep-lat",
    reportNumber = "MITP-19-021, MITP/19-021",
    doi = "10.1103/PhysRevD.100.014510",
    journal = "Phys. Rev. D",
    volume = "100",
    number = "1",
    pages = "014510",
    year = "2019"
}

@article{Borsanyi:2020mff,
    author = "Borsanyi, Sz. and others",
    title = "{Leading hadronic contribution to the muon magnetic moment from lattice QCD}",
    eprint = "2002.12347",
    archivePrefix = "arXiv",
    primaryClass = "hep-lat",
    doi = "10.1038/s41586-021-03418-1",
    journal = "Nature",
    volume = "593",
    number = "7857",
    pages = "51--55",
    year = "2021"
}

@article{Lehner:2020crt,
    author = "Lehner, Christoph and Meyer, Aaron S.",
    title = "{Consistency of hadronic vacuum polarization between lattice QCD and the R-ratio}",
    eprint = "2003.04177",
    archivePrefix = "arXiv",
    primaryClass = "hep-lat",
    doi = "10.1103/PhysRevD.101.074515",
    journal = "Phys. Rev. D",
    volume = "101",
    pages = "074515",
    year = "2020"
}

@article{Aubin:2022hgm,
    author = "Aubin, Christopher and Blum, Thomas and Golterman, Maarten and Peris, Santiago",
    title = "{Muon anomalous magnetic moment with staggered fermions: Is the lattice spacing small enough?}",
    eprint = "2204.12256",
    archivePrefix = "arXiv",
    primaryClass = "hep-lat",
    doi = "10.1103/PhysRevD.106.054503",
    journal = "Phys. Rev. D",
    volume = "106",
    number = "5",
    pages = "054503",
    year = "2022"
}

@article{Boccaletti:2024guq,
    author = "Boccaletti, A. and others",
    title = "{High precision calculation of the hadronic vacuum polarisation contribution to the muon anomaly}",
    eprint = "2407.10913",
    archivePrefix = "arXiv",
    primaryClass = "hep-lat",
    month = "7",
    year = "2024"
}

@article{Planck:2018vyg,
    author = "Aghanim, N. and others",
    collaboration = "Planck",
    title = "{Planck 2018 results. VI. Cosmological parameters}",
    eprint = "1807.06209",
    archivePrefix = "arXiv",
    primaryClass = "astro-ph.CO",
    doi = "10.1051/0004-6361/201833910",
    journal = "Astron. Astrophys.",
    volume = "641",
    pages = "A6",
    year = "2020",
    note = "[Erratum: Astron.Astrophys. 652, C4 (2021)]"
}

@article{Alguero:2023zol,
    author = "Alguero, G. and Belanger, G. and Boudjema, F. and Chakraborti, S. and Goudelis, A. and Kraml, S. and Mjallal, A. and Pukhov, A.",
    title = "{micrOMEGAs 6.0: N-component dark matter}",
    eprint = "2312.14894",
    archivePrefix = "arXiv",
    primaryClass = "hep-ph",
    doi = "10.1016/j.cpc.2024.109133",
    journal = "Comput. Phys. Commun.",
    volume = "299",
    pages = "109133",
    year = "2024"
}

@article{LHCHiggsCrossSectionWorkingGroup:2016ypw,
    author = "de Florian, D. and others",
    collaboration = "LHC Higgs Cross Section Working Group",
    title = "{Handbook of LHC Higgs Cross Sections: 4. Deciphering the Nature of the Higgs Sector}",
    eprint = "1610.07922",
    archivePrefix = "arXiv",
    primaryClass = "hep-ph",
    reportNumber = "CERN-2017-002-M, CERN-2017-002",
    doi = "10.23731/CYRM-2017-002",
    journal = "CERN Yellow Rep. Monogr.",
    volume = "2",
    pages = "1--869",
    year = "2017"
}

@article{Darricau:2025vcs,
    author = "Darricau, A. and Lee, H. and Orloff, J. and Teixeira, A. M.",
    title = "{Flavour and precision probes of a class of scotogenic models}",
    eprint = "2506.23383",
    archivePrefix = "arXiv",
    primaryClass = "hep-ph",
    doi = "10.1140/epjc/s10052-025-14946-9",
    journal = "Eur. Phys. J. C",
    volume = "85",
    number = "10",
    pages = "1234",
    year = "2025"
}

@article{Dubovyk:2018rlg,
    author = "Dubovyk, Ievgen and Freitas, Ayres and Gluza, Janusz and Riemann, Tord and Usovitsch, Johann",
    title = "{Complete electroweak two-loop corrections to Z boson production and decay}",
    eprint = "1804.10236",
    archivePrefix = "arXiv",
    primaryClass = "hep-ph",
    doi = "10.1016/j.physletb.2018.06.037",
    journal = "Phys. Lett. B",
    volume = "783",
    pages = "86--94",
    year = "2018"
}
\end{document}